\relax

\documentclass[letterpaper]{article} 
\usepackage{aaai20}  
\usepackage{times}  
\usepackage{helvet} 
\usepackage{courier}  
\usepackage[hyphens]{url}  
\usepackage{graphicx} 
\usepackage{graphicx}  
\usepackage{color}

\usepackage{verbatim}
\newcommand{\xxcomment}[4]{\textcolor{#1}{[$^{\textsc{#2}}_{\textsc{#3}}$ #4]}}
\newcommand{\hm}[1]{\xxcomment{magenta}{H}{M}{#1}}
\newcommand{\mn}[1]{\xxcomment{blue}{M}{N}{#1}}
\newcommand{\oa}[1]{\xxcomment{green}{O}{A}{#1}}

\newcommand{\orb}{ORB Descriptors}
\newcommand{\vgg}{VGG-16 CNN}
\newcommand{\siamese}{Custom-designed Siamese Neural Net}

\usepackage{subcaption}

\title{Dataset and Case Studies for Visual Near-Duplicates Detection in the Context of Social Media}
\author{Hana Matatov\textsuperscript{\rm 1}, Mor Naaman\textsuperscript{\rm 2}, Ofra Amir\textsuperscript{\rm 3}\\ 
\textsuperscript{\rm 1}Technion, {\rm hanama888@gmail.com}\\
\textsuperscript{\rm 2}Cornell Tech, {\rm mor@jacobs.cornell.edu}\\
\textsuperscript{\rm 3}Technion, {\rm oamir@technion.ac.il}
}

\begin{document}

\maketitle

\begin{abstract}
The massive spread of visual content through the web and social media poses both challenges and opportunities. 
Tracking visually-similar content is an important task for studying and analyzing social phenomena related to the spread of such content. 
In this paper, we address this need by building a dataset of social media images and evaluating visual near-duplicates retrieval methods based on image retrieval and several advanced visual feature extraction methods. 
We evaluate the methods using a large-scale dataset of images we crawl from social media and their manipulated versions we generated, presenting promising results in terms of recall.
We demonstrate the potential of this method in two case studies: one that shows the value of creating systems supporting manual content review, and another that demonstrates the usefulness of automatic large-scale data analysis. 
\end{abstract}

\section{Introduction}


The growing use of social media platforms has led to a large volume of visual content being distributed and spread across the globe~\cite{zannettou2018origins,highfield2016instagrammatics,paris2019deepfakes,dang2017offline}. Tools for accessing this content can allow researchers to study a variety of phenomena such as the spread of misinformation on the web and the distribution of content in different user groups. 
However, tracking of visual content on social media poses major technical challenges as it requires consistent data collection from social media content, as well as visual feature extraction and indexing, and image similarity matching methods that can work on a large-scale.
In this work, we aim to make multiple contributions to improve the tools available to social media researchers for studying and analyzing visual content. First, we build a large-scale training dataset of social media images. Second, we adapt and benchmark several existing computer vision techniques to find near-duplicates in this social media dataset. Finally, we design and share a pipeline that retrieves matching near-duplicate images based on a query image to allow large-scale studies in this domain. 
In this context, we consider exact-duplicates and near-duplicates, including substantial manipulations that are often distributed on social media (e.g. memes, captioned images, photoshopped images, rotated, cropped or spliced images, etc.)~\cite{paris2019deepfakes}.

Such dataset and retrieval tool can both enable large-scale automated analysis of social media content, and support manual processes for research and review. 
At scale, it can support research on the spread of visual content in different social media platforms~\cite{highfield2016instagrammatics,zannettou2018origins,dubey2018memesequencer,dang2017offline}.
In addition, such functionality can be the basis for user-facing systems. 
An example system can allow individuals and professionals such as journalists, to perform various important tasks in combating visual misinformation spread on social media in the context of news
~\cite{shu2017fake,matatovdejavu,gupta2013faking}. 


An ideal pipeline can update and maintain
a large dataset of images and index them in a way that allows efficient retrieval. 
The retrieval methods operating on such large-scale dataset need to be fast and sufficiently accurate to support the use cases listed above.
Matching and retrieval of near-duplicates in the context of social media is not a trivial computer vision task, as it needs to be robust to varied manipulations of the image, including edits which may be adversarial to avoid detection~\cite{marwick2017media,marra2018detection,marra2019gans,chesney2019deep}. 
Images of interest may also vary according to many other aspects including image quality, image source and type of content. 

To address these challenges, we develop a dataset of social media images and manipulations. We build an indexing and retrieval pipeline for the task, adapting and benchmarking three existing methods for extracting image visual features~\cite{rublee2011orb,simonyan2014very,bromley1994signature}. 
The core dataset we use is of $2,290,683$ images sourced from Reddit and 4chan. 
We use a sample of $64,957$ images, as well as manipulated versions of a subset of these images, to benchmark the performance of the  near-duplicates retrieval process.
Evaluation results show that for a wide range of manipulation types, our process obtains high recall for matching images. 
The dataset will be made available to the research community for further studies of visual content distribution through social media.

To demonstrate the potential use of the developed methods, we explore two case studies using a dataset of $631,018$ images we crawled from Reddit. 
The first case study examines in detail the performance of the retrieval methods on queries with generated manipulated versions of the original image. The case study results show that the developed methods achieve high recall values even on a large-scale index.
The second case study examines the potential of using our methods for tracking the spread of matching images across different social media platforms.
Using an automated process which determines whether an image is a near-duplicate of a given query image, we show 
how the system can be useful for tracking images distributed across Twitter, 4chan and Reddit. 

\section{Related work}
Our work builds on prior research on near-duplicates detection and retrieval, in order to support a growing body of research in social media \textit{visual} analytics. 
Research on distribution of fake news and misinformation --- to use an example of social media content that is often distributed via repeated and manipulated images --- had focused on textual content~\cite{vosoughi2018spread,grinberg2019fake,guess2019less,allcott2017social}.  
Computational techniques for fake news detection had also predominantly focused on text~\cite{shu2017fake}.

At the same time, there is a growing interest in analysis of visual content~\cite{highfield2016instagrammatics,shu2017fake,paris2019deepfakes}, often using similarity-based techniques. Several studies focused specifically on Memes~\cite{dubey2018memesequencer,zannettou2018origins,dang2017offline}.
Zannettou et al.~\shortcite{zannettou2018origins} 
collected images from four Web communities (e.g. Reddit and Twitter). They used a processing pipeline based on Perceptual Hashing~\cite{monga2006perceptual} and clustering techniques to assess the popularity and diversity of memes in the context of each community, model the interplay between the communities and quantify their reciprocal influence. 

These projects, studying the spread of visual content on social media, demonstrated the existing gaps and the need for a tool for tracking the spread on social media of a \textit{wide} range of near-duplicates in a large-scale and real-time manner. 
The near-duplicates may be exact-duplicates, or any manipulated versions that often appear on social media, including extreme or adversarial manipulations (e.g. memes, sever cropping, splicing etc.).
Our work aims to address these gaps and provide a dataset and benchmarks specifically in the context of visual near-duplicates on social media.

In computer vision, several works have addressed the problem of near-duplicates detection, some of them using image indexing and retrieval for this purpose~\cite{dong2012high,ke2004efficient,an2017near,chum2007scalable,zhu2008near}.
Retrieval of near-duplicate images had frequently build on SIFT-based~\cite{lowe2004distinctive} techniques~\cite{ke2004pca,dong2012high}. For example, in \cite{dong2012high}, the authors use SIFT features representation, examine different similarity matching approaches and present benchmarking in terms of memory and computation time.
An et al.~\shortcite{an2017near} compared local and global feature representations, including several existing Neural Network architectures. They retrieve near-duplicates using clustering and Binary Hashing~\cite{he2013k}. 
Systems like Google Reverse Image Search (RIS)\footnote{https://images.google.com/}~\cite{barroso2003web} provide functionality for finding duplicates, but are not specifically trained and optimized in the context of social media platforms. 
For example, RIS only supports retrieval of images that still exist at the time of searching.
Therefore, coverage of social content is incomplete due to 
the fact that some of the social media sources are ephemeral.
In addition, Google RIS performs both visual and semantic matching (which is beyond visual matching for near-duplicates detection), and does not export metadata like match score to make it usable for automated systems.

Other studies were done in our context, and benchmarked their methods by using, among others, Web or social media datasets~\cite{moreira2018image,pinto2017provenance}.
For example, Moreira et al.~\shortcite{moreira2018image}
investigated the \textit{image provenance analysis} task, i.e., retrieving from a large index the set of original images whose content presented in the query image, as well as the detailed sequences of transformations that yield the query image. 
Similarly, Pinto et al.~\shortcite{pinto2017provenance} 
developed provenance filtering methods  
and conducted experiments using the NIST (National Institute of Standards and Technology) dataset, which was developed for the Media Forensics Challenge Evaluation\footnote{\url{https://www.nist.gov/itl/iad/mig/media-forensics-challenge}} and focuses, among other tasks, on provenance filtering. 

We build on the near-duplicates retrieval research and ideas and use a similar approach in the techniques we benchmark here.
Our work also uses image datasets and types of manipulations more focused on the social media context. 
In particular, we use a broader definition of \textit{``near-duplicate''} images, due to our visual tracking on social media context.
I.e., we consider an image a near-duplicate even if an extreme manipulation, or a combination of several manipulations were made.
Thus, in the following sections, we present a benchmark of the near-duplicates detection methods on our social media dataset.
Namely, in this context, our contribution is the comparison and combination of several near-duplicates detection advanced techniques, to optimize the specific task of visual near-duplicates detection in the context of social media.

While there are also methods for pairwise evaluation of image similarity, they are not feasible in the context of fast retrieval of matching images from a large-scale dataset. 
Some address the problem of detecting \textit{whether} an image is photoshopped or forged~\cite{lago2019visual,lago2018image,marra2018detection},
spliced~\cite{huh2018fighting}, or generated by adversarial networks~\cite{marra2018detection,marra2019gans,xuan2019generalization,mccloskey2018detecting,zhang2019detecting}.
These tasks are complementary to the \textit{retrieval} task we address. 
The problem we address is different, in that we do not aim to determine \textit{whether} an image has been manipulated, but rather retrieve additional appearances of the image or its near-duplicates from a large-scale dataset.

\section{Visual Feature Extraction}
\label{sec:feature_extraction}
Retrieval of images from a large-scale dataset requires methods for  representing the images such that they can be efficiently indexed and stored.
To this end, we examined several existing methods for image representation and key-points extraction~\cite{radenovic2018revisiting}.
Ultimately, we used three different feature representations, based on known techniques but applied specifically in this context, i.e., such that they would support retrieving exact matches or manipulated versions of an image (as opposed to semantically similar images).
One can consider these existing techniques as three baseline approaches, as we are comparing the performance of these known approaches in our context (Section~\ref{sec:eval}).
In the following, we describe the three approaches in detail.

Note that we considered using Perceptual Hashing (pHash)~\cite{monga2006perceptual}, a process of constructing a hash value that uniquely identifies an input image based on the contents of the image. 
However, even minor edits on images result in different pHash values, thus in its basic form it is more suitable for exact-duplicates detection.
In order to use Perceptual Hashing for near-duplicates detection, a similarity measurement between the pHash values and a threshold should be defined~\cite{monga2006perceptual}.  
In this work, we aim to detect visually-similar images, even if they were subjected to quite extreme manipulations.
Hence, we did not use pHash. 

\subsection{Computing ORB Local Descriptors}
\label{sec:orb}
We explored several local descriptors computation methods\footnote{Such as HOG~\cite{dalal2005histograms}, SIFT~\cite{lowe2004distinctive} and ORB~\cite{rublee2011orb}.}.
Ultimately, we used ORB descriptors~\cite{rublee2011orb}, which are 256 bit binary vector describing unique selected points of interest in the image.
We limited the number of descriptors for each image to be its 200 most significant descriptors (at most) as it resulted in improved performance~\cite{rublee2011orb}.
We chose to use ORB descriptors since they are based on an orientation component, therefore they are rotation invariant (in-plane rotation) and quite resistant to noise. Hence, ORB descriptors are likely to remain in images even after transformations and manipulations~\cite{rublee2011orb}. 
We used Principal Component Analysis (PCA)~\cite{abdi2010principal} 
to reduce each feature to a vector of 128 bits. The PCA model was trained by bootstrapping using 70,000 of the images.

\subsection{Using Pre-trained VGG-16 CNN}
\label{sec:vgg}
The second method for feature extraction uses a pre-trained Convolutional Neural Network (CNN) called VGG-16~\cite{simonyan2014very}.
VGG-16 is a 16-weight-layers CNN which was trained to classify images from the ``ImageNet'' dataset. 
We chose to examine VGG-based features since CNNs have been shown to be useful for many computer vision tasks\footnote{We explored some other Neural Network  architectures, which achieved similar results, thus we used VGG-16.}. 

Since we are interested in extracting image features and not in classification, we used the output of the last max pooling layer rather than the final output of the network (ignoring the last three fully-connected layers).
Babenko et al.~\shortcite{babenko2014neural} 
and Sermanet et al.~\shortcite{sermanet2013overfeat} showed that features emerging in the upper layers of a convolution neural network (CNN) trained for classification can serve as good high-level descriptors of the visual content of the image.
While some prior works use the output of one of the two following fully-connected layers (which appear before the final fully-connected classification layer)~\cite{yang2017yum}, we chose to stop at an earlier layer as such feature representation has been shown to enable better generalization to images that are different than those that appear on the ``ImageNet'' dataset~\cite{guerin2017cnn}. 
The last max pooling layer’s output is a 7X7X512 matrix.
We then apply L2 normalization on each 512 length vector.

\subsection{Using Trained Custom-designed Siamese Neural Network}
\label{sec:siamese}
Our third approach tried to optimize the feature representation specifically for the task of retrieving similar images. We designed a neural network for this task, inspired by the Siamese Neural Network architecture~\cite{bromley1994signature}. Siamese architectures typically consist of two identical neural networks with shared weights, where each network takes an input sample and outputs a feature vector. Then, a distance measure is calculated between these vectors, which is used to classify whether the inputs are of the same class or not. 
Thus, the architecture of the model consists of two essential parts - feature extraction and similarity measurement. After training the Siamese model weights, we used the output of the first part for feature extraction. 

We used the  VGG-16 network architecture for feature extraction, only until the last max pooling layer, as described in the previous subsection. 
However, this time we fine-tuned the network weights by activating the learning in the last two layers (while freezing all layers until that, i.e. used the pre-trained weights).
This network outputs a 7X7X512 dimensions feature matrix.
We added a Global Max Pooling layer that takes the maximal value among the 49  for each of the 512 entries, resulting in a 1X512 feature matrix.

\begin{figure}[h]
\centering
\includegraphics[width=1.1\columnwidth]{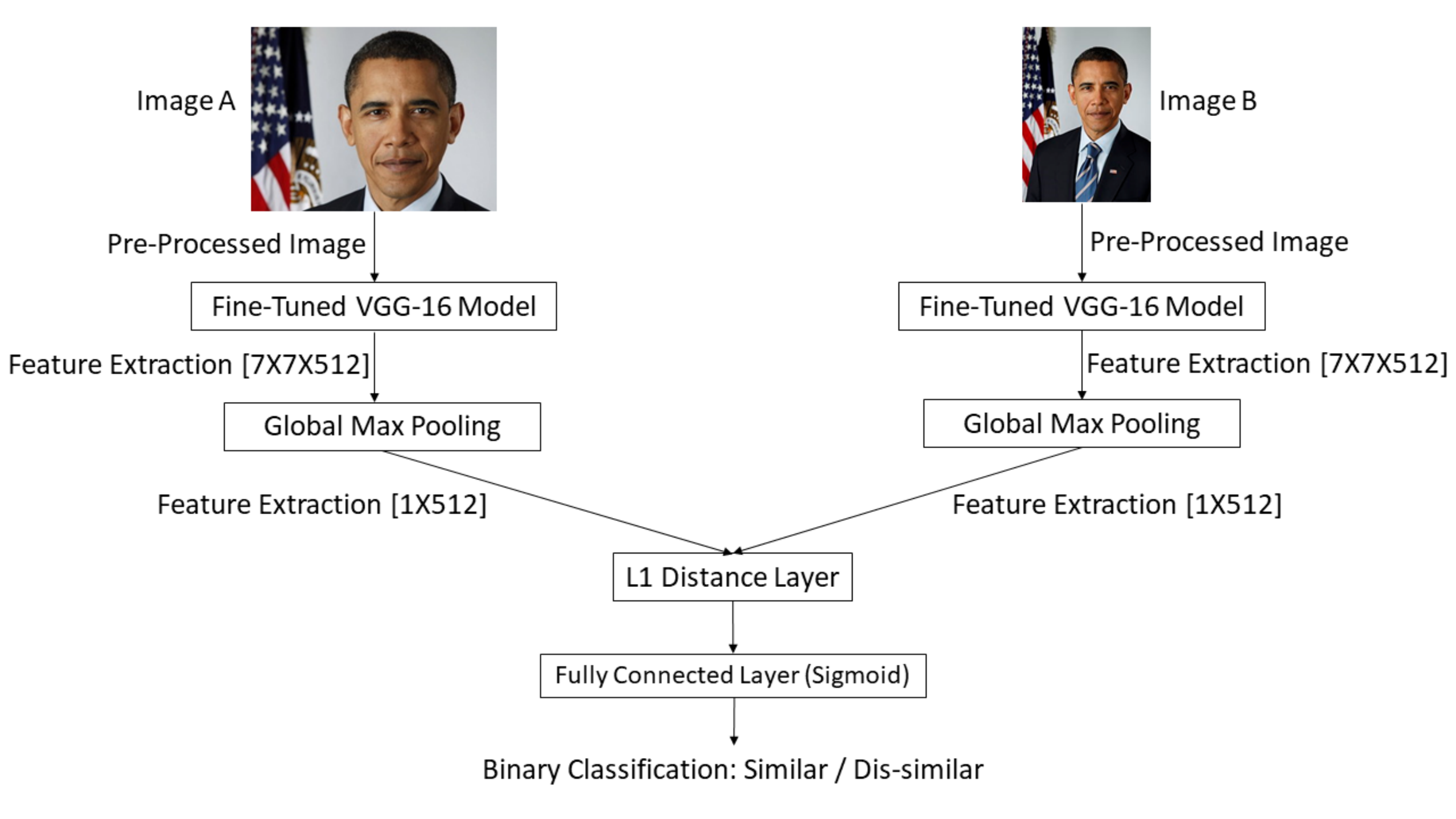}
\vspace{-0.5cm}
\caption{Custom-designed Siamese Neural Network Architecture. 
}
\label{fig:Architecture}
\end{figure}

The similarity measurement component of the  network  received as input two extracted feature vectors (each 1X512), calculated for two images as explained above.
The first layer in this component is an L1 distance layer between the two feature vectors.
It is followed by a fully connected layer with Sigmoid activation, which outputs the classification prediction - whether the two input images are classified as similar or dis-similar. 
Figure~\ref{fig:Architecture} shows the architecture of the \siamese, which consists of two identical feature extraction networks with shared weights, and a similarity measurement component.

In order to train the \siamese, we used $40,595$ images from the Breaking News dataset\footnote{\url{http://www.iri.upc.edu/groups/perception/#BreakingNews}}, which includes images from news articles from several major newspapers and media agencies. The images collected during 2014, covering  a wide variety of topics, including politics, local news, sports, healthcare etc.~\cite{ramisa2017breakingnews}.

We needed labeled data - similar and dis-similar image pairs. To this end, we generated $81,190$ similar pairs and $81,200$ dis-similar pairs, a total of $162,390$ pairs of images.
For similar image pairs we used manipulations of the same source image (or the source image itself).
For each source image, we randomly generated two pairs which are consisting of different combinations of versions of the source image. These were randomly sampled from the 23 versions (the source image and its 22 manipulations).

Dis-similar image pairs were created from versions of different source images. To make training more effective, we chose the pairs in an adversarial manner, by selecting images that are relatively visually similar, although they did not originate from the same source image. To identify such pairs, we clustered the images to 100 clusters using K-Means~\cite{lloyd1982least,kanungo2002efficient}. The image clustering based on feature representation from a pre-trained VGG-16 neural network (the second Fully Connected layer, which outputs a 1X4096 dimensions vector)~\cite{guerin2017cnn}. 
As a sanity check of the clustering approach, we sampled pairs of images from each cluster and manually determined whether they were visually similar (e.g., two images of a football field would be considered similar). We obtained 84\% accuracy, which was satisfactory for our task of creating visually similar (yet dis-similar) image pairs.
We randomly sampled 812 pairs of images from each cluster to use in the training.

We split, separately, the similar and dis-similar image pairs to train, validation and test sets (70\%, 15\% and 15\% respectively). 
We used Adam optimizer with a learning rate of 0.001 and Binary Cross-Entropy loss, since we trained to minimize loss for distinguishing between similar and dis-similar pairs of images.
We trained the network with batches of size 2048 for 16 epochs.
Overall, both train and validation accuracy scores improved during training, achieving a maximal score in the last epoch (epoch 16). The gap between final train accuracy (=99.04\%) to final validation accuracy (=97.41\%) was not very large, suggesting that the model does not severely suffer from overfitting. We achieved 97.27\% accuracy on the test data.

Finally, after having a trained \siamese, we used it as a model for feature extraction for image indexing and retrieval. This was done by feeding an image to the model and saving the output of the Global Max Pooling layer as a feature vector of 1X512 dimensions. We normalized each 512 length vector using L2 normalization. 

\section{Image Indexing and Retrieval}
\label{sec:retrieval}
We aimed to develop a method for retrieving the images that are most visually similar  to a given query image, supporting the task of near-duplicates detection.
The retrieval process is expected to be real-time, so that potential users could rely on it to query new images, e.g. during a breaking news event.
Thus, we could not use unsupervised clustering methods, which are characterized by a long ``online'' phase~\cite{lloyd1982least,vattani2011k}, or standard pairwise similarity computations, according to distance metrics (e.g. Manhattan or Euclidean distance),which would also be too slow.

The retrieval  process we developed consists of two main phases.
In the ``offline'' phase, we extract features for all the images in the dataset (as explained in Section~\ref{sec:feature_extraction}), and index them.
In the ``online'' phase, given an incoming query image, we invoke the similarity matching process for retrieving the most similar images in the index.
In order to allow efficient image search, such that the ``offline'' phase requires time but the ``online'' phase is fast, we constructed the index using the FAISS (Facebook AI Similarity Search) package, a library for efficient similarity search and clustering of dense vectors~\cite{johnson2019billion}.
Specifically, we used ``IndexFlatL2'' which is characterized by exact search results. 
Moreover, FAISS provides an implementation for fast k-nearest-neighbor search (KNN)~\cite{ge2013optimized}, using squared Euclidean distance as its distance metric.
We extracted the features of all existing images in the dataset using the three methods described in the previous section, and created three indexes accordingly.

We used KNN with k=100 to obtain similarity scores and retrieve from the index the most similar images to the query image.
Hence, we implemented three pipelines for image retrieval, using the three different feature extraction methods.
In all three cases, we first extracted features for the query image using the corresponding feature extraction method. We then search for the most similar $N$ images in the index. When ORB Descriptors or VGG-16 CNN are used, we first find the $k$-nearest-neighbors in the index for each feature of the query image. The similarity score between the query image and each of the images in the index is calculated as the number of its features matched by the KNN search to the query image features. When using features extracted from the \siamese, the KNN similarity scores are simply the squared Euclidean distances between the feature vector of the query image to those of the images in the index, since this feature extraction methods results in 1-dimensional vector per image. 

\section{Dataset}
\label{dataset}
In this section we describe the dataset we collected and used to benchmark the process of indexing and retrieving matching, exact or near-duplicate, social media images for a query image.
For these images, we further automatically generated a variety of manipulation types, to enable evaluation of the robustness of retrieval methods. 

\subsubsection{Images Crawled from Social Media.}
We  especially focus on Reddit
and 4chan, as these channels are known for spreading information without regulation, including spreading fabricated and photoshopped content~\cite{zannettou2018origins,dang2017offline,moreira2018image}. 
We crawled two subreddits that are known as sources of misleading content and image violations:  ``r/the\_donald'' and  ``r/conspiracy''.
We focus on images from posts that appear in the ``/pol/ - Politically Incorrect'' board on 4chan.
We crawled the image content from the subreddits and  4chan starting from July 2018. 
At the time of evaluation (March $2020$), the dataset included $2,290,683$ images: $1,659,665$ from 4chan and $631,018$ from Reddit.
We will make the dataset available to the research community upon publication of the paper.

\subsubsection{Manipulations Generation.}
\begin{figure}
\centering
\begin{tabular}{cc} 
\subcaptionbox{Source Image\label{a}}
{\includegraphics[width = 1.4in, height = 1.2in]{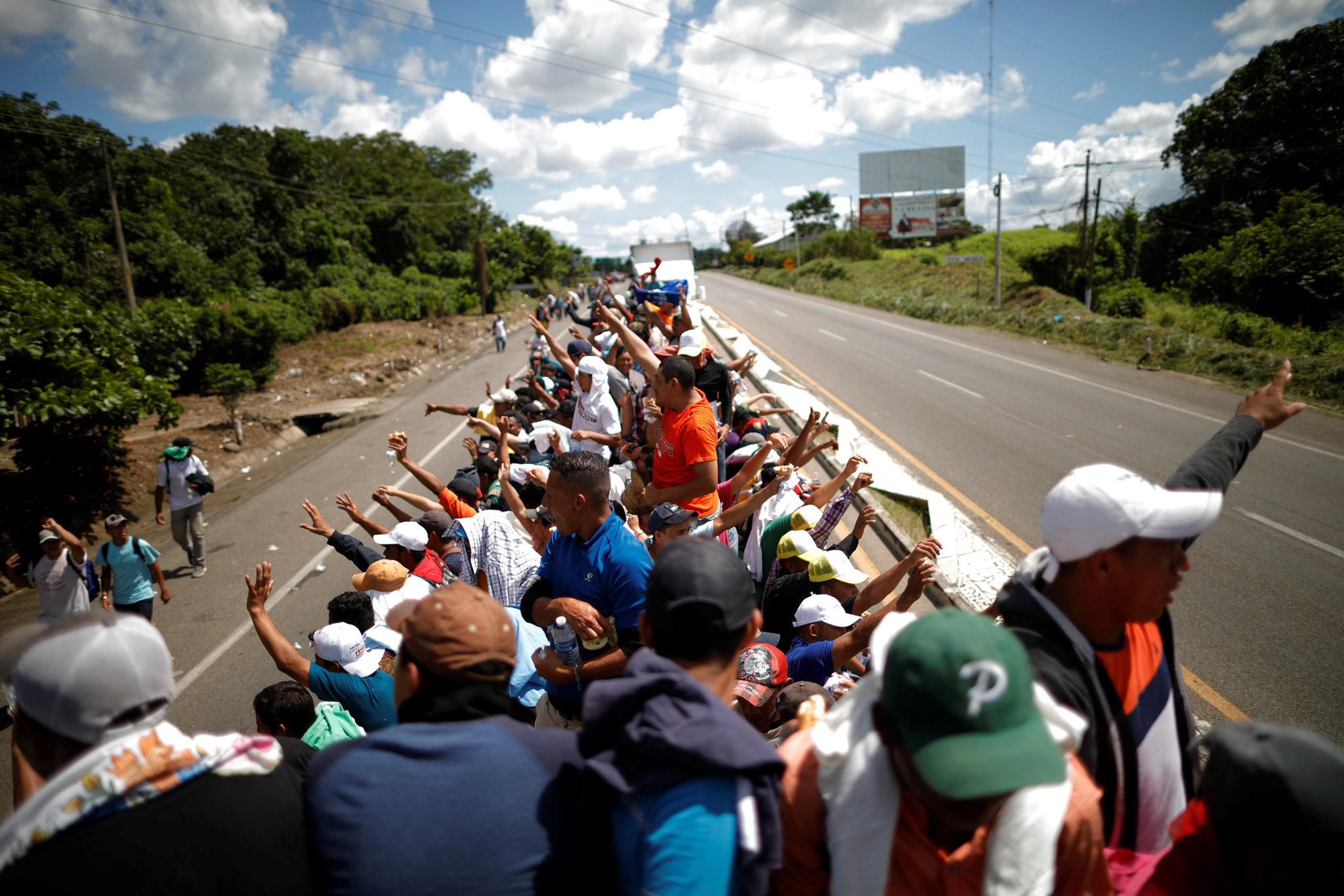}} &
\subcaptionbox{Gaussian Noise (SD=8)\label{b}}
{\includegraphics[width = 1.4in, height = 1.2in]{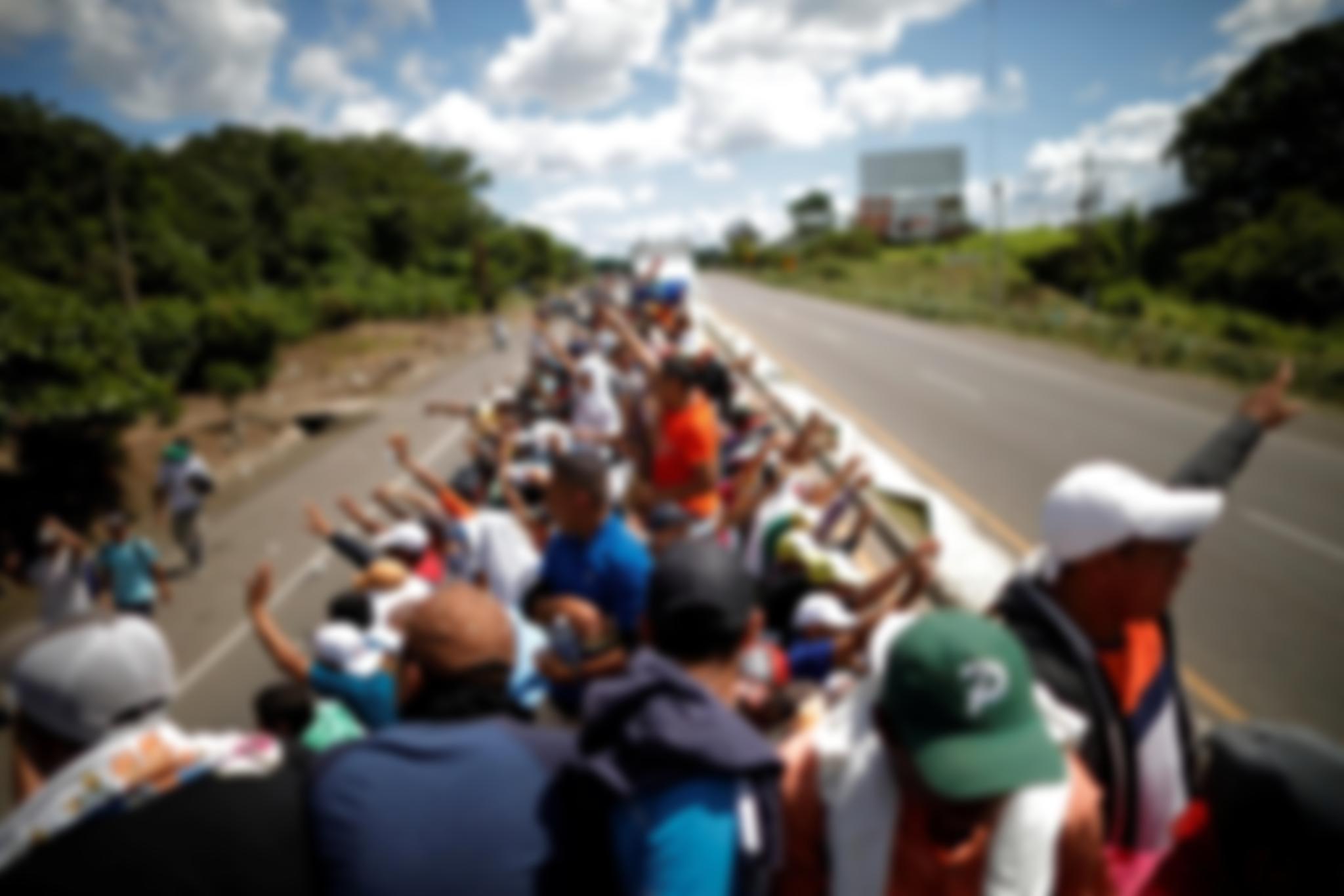}} \\
\subcaptionbox{Crop from (1/3,1/3)\label{c}}{\includegraphics[width = 1.4in, height = 1.2in]{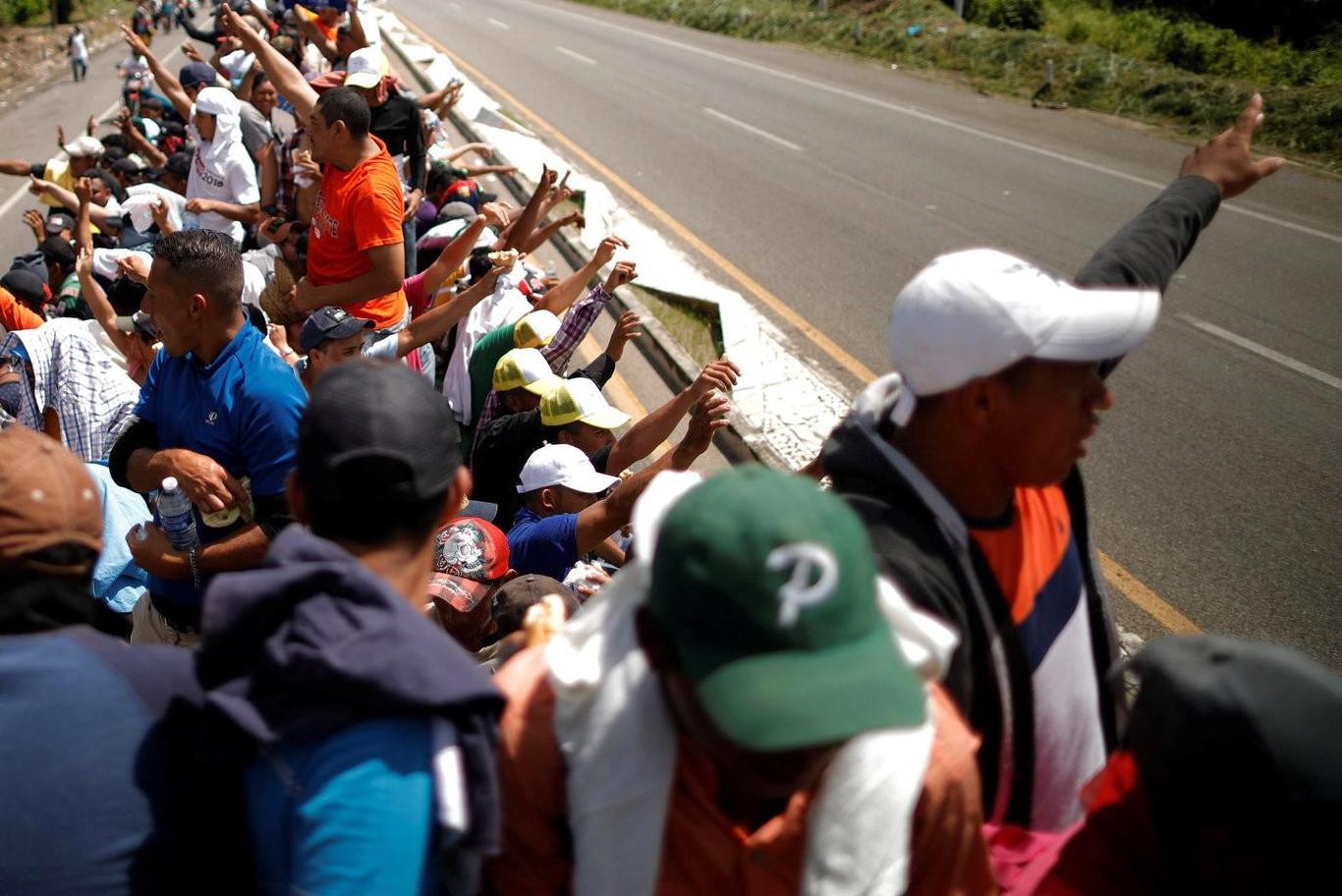}}&
\subcaptionbox{Black\&White Color Transformation\label{d}}{\includegraphics[width = 1.4in, height = 1.2in]{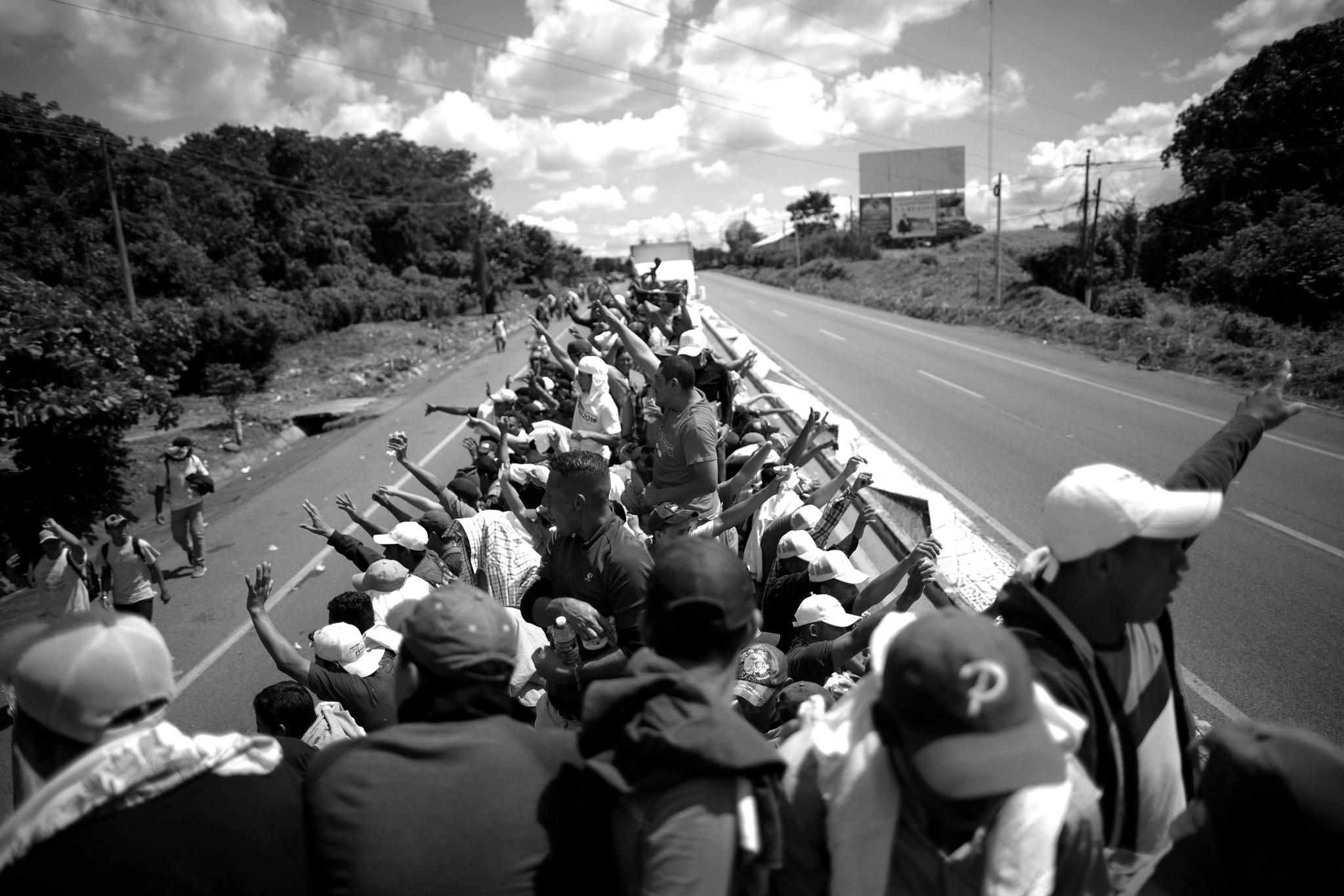}} \\
\end{tabular}
\vspace{-0.1cm}
\caption{Example of the generated manipulations.
}
\label{visual_example_manipulations}
\end{figure}

In order to benchmark the similar images retrieval process, we had to create a ground truth collection, which contains source images and their manipulations. 
We wanted to generate a set of manipulations that mimic common scenarios of image modifications that occur on social media, where images are frequently down-sampled, cropped, spliced etc.
In this step, we generated the manipulations using available tools for automatic image manipulations generation.
Specifically, given an original source image, we automatically generated 22 consistent manipulations, including:
\begin{itemize}
\item Adding Gaussian noise with different standard deviation parameters - 2, 4 or 8.
\item Cropping different areas of the image - leaving only the bottom-right quarter, the bottom-right two thirds or the top-left two thirds.
\item Horizontal flipping of the image.
\item Rotation transformation - rotating the image 5 or 10 degrees clockwise and 5 or 10 degrees counter-clockwise.
\item Changing resolution - resizing the image to 20\%, 40\% or 80\% of its size.
\item Color transformations - changing the color channels from RGB to GBR.
\item Convert image to grey level scale (black\&white image).
\item Adding simple text on top of the image.
\item Adding different visual markups 
on top of the image.
\item Filtering with ``motion'' filter (approximate  the linear motion of the camera). Motion with different lengths: 10, 15, 20 and different angle degrees (counter-clockwise): 15, 20, 25, respectively.
\end{itemize} 
Figure~\ref{visual_example_manipulations} shows an example of one source image from our dataset
and some of its manipulations.
Further, on Section~\ref{case_study_b}, we present a case study analysis regarding matching images that were not generated by us, rather collected by us from several social media sources.

\section{Evaluation Procedure}
\label{sec:eval}
Our overarching goal is creating an image retrieval pipeline for social media that can operate at scale. The basic capability needed for such a pipeline is effective retrieval of near-duplicate images. The retrieval performance is affected by the visual features that build the index. Therefore, the first step in our evaluation aims to test the effectiveness of the existing feature extraction methods for this task.


For this step we used a limited subset of images and queries.
In Section~\ref{case_study_a}, we use a larger dataset to provide a more detailed analysis of the impact of the different types of manipulations on retrieval performance.

To create the index, we randomly sampled $64,957$ images from the dataset.
To generate a set of query images, we performed a multi-step process. First, we randomly chose a subset of $10$ images from the $64,957$ images. 
Second, for images included in this subset, we generated a set of manipulations.
This process resulted in $221$ images, including the $10$ source images and all their manipulations.
We used these $221$ images as query images, such that we know their source images exist in the index.

For each query image, we examined the ranking of its original source image (from which the manipulation was generated) in the retrieved results. 
We calculated  recall at $1$, $3$ and $10$ (denoted $recall@1$, $recall@3$ and $recall@10$, respectively). $recall@1 = 1$ if the source image was the top ranked retrieved result, $recall@3 = 1$ if the source image was ranked in the top $3$ results and $recall@10 = 1$ if the source image was ranked in the top $10$ results.
Note that this applies for only one source image for each query image, even though the query images may have other near-duplicates in the index.

\section{Results}
\label{sec:results}
Overall, the three methods we used -- \orb,~\vgg~and~\siamese -- were successful in retrieving images, obtaining high recall values.
A summary of the results is shown in Table~\ref{results_table}.

If the source image was successfully retrieved, it often happened in the top three results.
Table~\ref{results_table} shows that the results for each method improved significantly as the number of retrieved images grows from one
($recall@1$) to three ($recall@3$).
A possible explanation is that frequently the source image and other existing near-duplicates retrieved with the same similarity score, and randomly retrieved as the first or second result. 
We do not observe a significant gain between the average $recall@3$ and $recall@10$. That is, if the similar image was not retrieved among the first three results, it was not likely part of the top 10 results, either.

VGG16-based image retrieval obtained the best performance in terms of $recall@3$ and $recall@10$, ($recall@3, recall@10 >= 0.923$), while the ORB-based and Siamese techniques achieved lower performance ($recall@3, recall@10 <= 0.882$). 
The differences between $recall@3$ and $recall@10$ values obtained by the VGG16-based and Siamese feature extraction methods are statistically significant ($p$-value = 0.017 and $p$-value = 0.044 respectively).
The difference between the $recall@10$ values obtained by the VGG16-based and ORB-based methods are also statistically significant ($p$-value = 0.044)\footnote{For each value ($recall@3$ and $recall@10$) we conducted statistical significance tests by using three Chi-square tests (for each pair of feature extraction methods).}.

\begin{table}[]
\centering
\small
\addtolength{\tabcolsep}{-1pt}
\begin{tabular}{|c|c|c|c|}
\hline
                                                                              & \textbf{R@1}                                                     & \textbf{R@3}                                                     & \textbf{R@10}                                                    \\ \hline
\textbf{ORB Descriptors}                                                      & \begin{tabular}[c]{@{}c@{}}0.706\\ {[}0.65, 0.76{]}\end{tabular} & \begin{tabular}[c]{@{}c@{}}0.864\\ {[}0.82, 0.91{]}\end{tabular} & \begin{tabular}[c]{@{}c@{}}0.882\\ {[}0.84, 0.92{]}\end{tabular} \\ \hline
\textbf{VGG-16 CNN}                                                           & \begin{tabular}[c]{@{}c@{}}0.557\\ {[}0.49, 0.62{]}\end{tabular} & \begin{tabular}[c]{@{}c@{}}0.923\\ {[}0.89, 0.95{]}\end{tabular} & \begin{tabular}[c]{@{}c@{}}0.941\\ {[}0.91, 0.97{]}\end{tabular} \\ \hline
\textbf{\begin{tabular}[c]{@{}c@{}}Custom-designed\\ Siamese NN\end{tabular}} & \begin{tabular}[c]{@{}c@{}}0.629\\ {[}0.57, 0.69{]}\end{tabular} & \begin{tabular}[c]{@{}c@{}}0.846\\ {[}0.80, 0.89{]}\end{tabular} & \begin{tabular}[c]{@{}c@{}}0.882\\ {[}0.84, 0.92{]}\end{tabular} \\ \hline
\end{tabular}
\caption{Average recall (and $95\%$ confidence intervals) for different top K values (K=1,3,10) for each of the methods (\orb, \vgg, \siamese) when using the social media index.}
\label{results_table}
\end{table}

In the next section, we focus on the VGG-16 method to provide a more detailed analysis of the impact of the different types of manipulations on retrieval performance.
Note that VGG-16 also satisfies the requirement to provide a real-time pipeline. On average, the current non-parallel performance requires $1.3$ seconds to extract the features of the query image, and additional $3.9$ seconds to search and retrieve its most visually similar images in the index. These times can be improved by using a different index type, which is optimized for efficiency (the FAISS package supports indexes optimized for performance, which we are not using here), and by using parallelism.

\section{Retrieval Case Studies}
Following the evaluation of the image indexing and retrieval process, 
we examine its usefulness and potential in two realistic case studies 
demonstrating the potential applications of the approach to tracking of visual content distribution in social media.

Both case studies use an expanded index that includes $631,018$ images we crawled 
from the political subreddits listed in Section~\ref{dataset}. 
For the case studies, we used the same image indexing and retrieval process as described in previous sections (\ref{sec:feature_extraction} and~\ref{sec:retrieval}), extracting image features using the \vgg~method for the visual feature extraction step.

\subsection{Case Study 1: Finding Manipulated Versions on Reddit}
\label{case_study_a}
The goal of this case study is to examine a scenario where a user wishes to check whether manipulated versions of an image have appeared on the web.
We simulated this scenario by using a large index of images from Reddit, and querying manipulated versions of a subset of those images. This way, we know the index includes their source image.
This case study is similar to the evaluation procedure presented before (Section~\ref{sec:eval}), but here we use only \vgg~for feature extraction (as it obtained the best performance), and query an index which is larger in size, making it more realistic in the context of retrieval from social media images ($631,018$ versus $64,957$ images).

\subsubsection{Procedure.} 
We randomly selected $554$ images from the $631,018$ crawled Reddit images, to serve as source images.
Then, we created visual manipulations for these images, according to the procedure described in Section~\ref{dataset}.  
The source images together with their manipulated versions resulted in $12,726$ query images. 

For each query image, we queried the new index and examined the ranking of its original source image in the retrieved results. 
Similarly to previous analysis, we calculated $recall@1$, $recall@3$ and $recall@10$ for each query. 

\subsubsection{Results.}
Overall, our indexing and retrieval process was successful in retrieving similar images, with $recall@1$, $recall@3$, $recall@10$ values of $0.662$, $0.822$, $0.881$, respectively.
Although we retrieved from a 10-fold larger index than the initial evaluation, the recall values remain high. 
For example, in the evaluation above, we had $recall@10 = 0.941$ when using \vgg, whereas now $recall@10 = 0.881$. 

\begin{figure}[h]
\centering
\includegraphics[width=1.1\columnwidth]{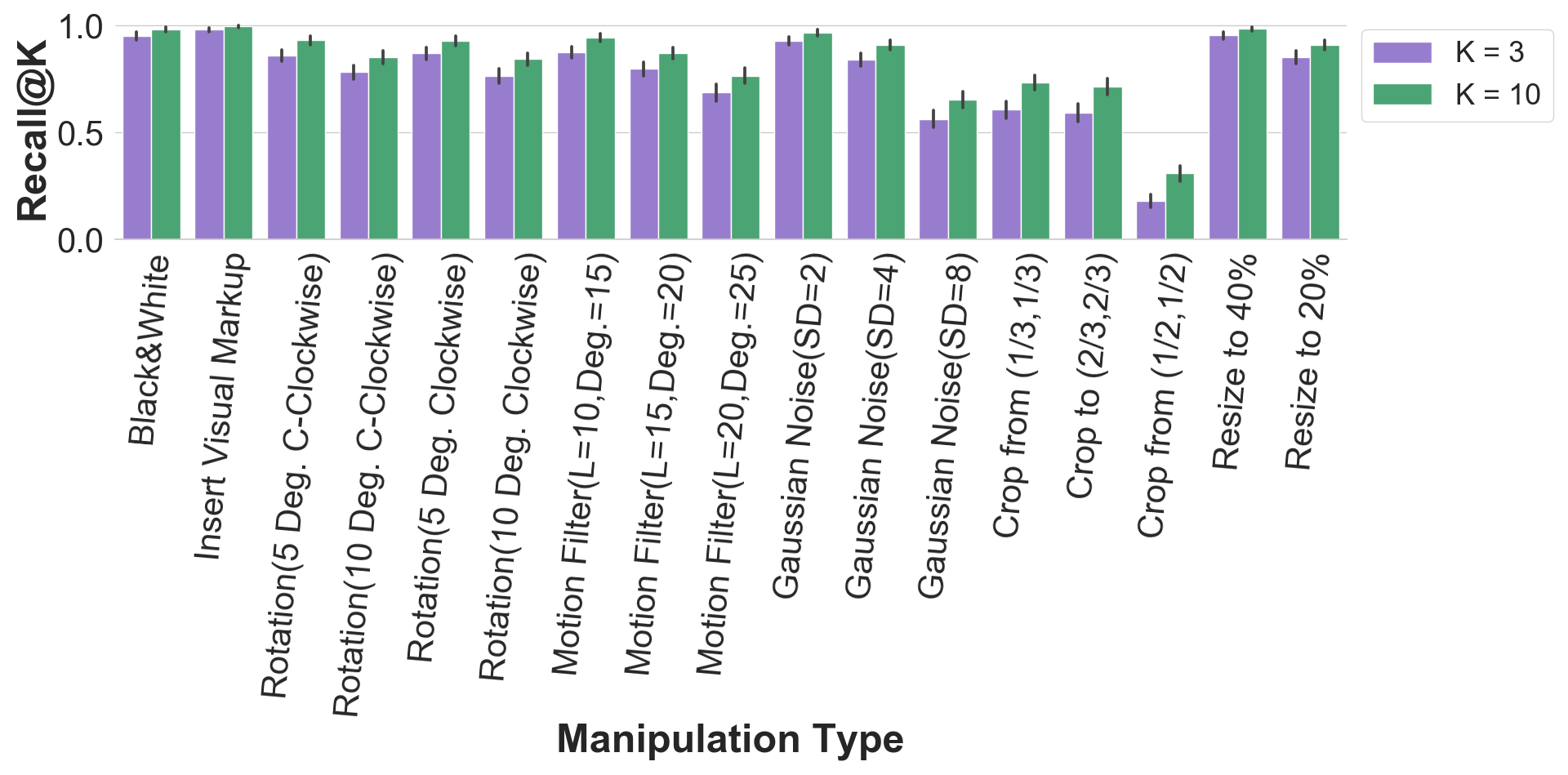}
\vspace{-0.55cm}
\caption{$Recall@3$ and $Recall@10$ for different manipulations (case study 1). 
}
\label{fig:recall_3_10_vgg_reddit}
\end{figure}

Performance varied depending on the type of manipulation used.
Figure~\ref{fig:recall_3_10_vgg_reddit} shows the average $recall@3$ and $recall@10$ values obtained for each of the manipulation types\footnote{Due to space considerations, we do not show some of the manipulation types for which  high recall values were obtained and are similar manipulations to other manipulations in the figure (e.g. resize to $80\%$ or GBR color transformation).}.
As shown in Figure~\ref{fig:recall_3_10_vgg_reddit}, for most manipulation types we generated, the source image was retrieved among the top three results.
However, some types of manipulations resulted in lower recall values. 
In particular, the most extreme addition of Gaussian noise (SD=8) resulted in relatively low recall values, as well as significant cropping actions.

The case study results show that our similarity matching and retrieval process is useful for real-world use cases.
Also, results indicate that the process is considerably robust to a wide range of images and manipulation types.


\subsection{Case Study 2: Finding Earlier Appearances of Similar Images}
\label{case_study_b}
The goal of this case study is to examine the usefulness of our process for large-scale (hence, automated) measurement of the spread of matching images between different social media platforms.
In particular, we study whether a specific set of images, posted in political context on Twitter or on 4chan, first appeared --- as exact or near duplicates --- on the political subreddits we indexed. 
Thus, this case study also provides an evaluation of the real-world applicability of our pipeline and dataset.

The full automation required by this large-scale use case requires an additional computational step where a system automatically decides whether the top-ranked matches can indeed be considered near-duplicates. We describe this additional automation here, and report on the results below.

In the initial evaluation  
we verified that, with high accuracy, our process is successful in retrieving the most similar images that exist in the index (Section~\ref{sec:results}).
But, if the index does \textit{not} contain near-duplicates, the retrieved results will not be relevant.
We devised a bottleneck-based procedure, where we retrieve the most similar images using our index, and then use machine learning to decide whether  these images are near-duplicates of the original query image.


To this end, we compute the Perceptual Hashing (pHash) values~\cite{monga2006perceptual} for the match between the query image and a retrieved result, as well as the retrieval similarity score returned from our index, and train a machine learning model to decide whether the images are indeed near-duplicates. 

\subsubsection{Training.} 
To train the model, we used a sample of $240$ images from Twitter and 4chan. For each image, we retrieved its most similar image from our expanded index. 
We then visually examine the retrieved results and manually assess whether the image is a duplicate (or near-duplicate) of the query image. 
A pair of query and retrieved image are considered as matching even if they were subject to various types of manipulations, including challenging manipulations such as extreme cropping, low resolution, photoshopped images, memes and combination of several manipulation types.
These manipulated versions were created and distributed by social media users, without any constrains.
For this case study, we only annotated the first retrieved result for each query to create our training data, but we tested on the first four retrieved results (as we note below).



\begin{figure}
\centering
\begin{tabular}{cc} 
\subcaptionbox{4chan Query\label{a}}
{\includegraphics[width = 1.2in, height = 1.4in]{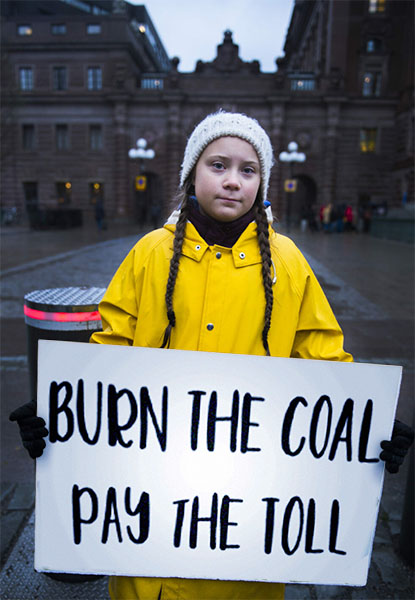}} &
\subcaptionbox{Reddit Retrieval\label{b}}{\includegraphics[width = 1.2in, height = 1.4in]{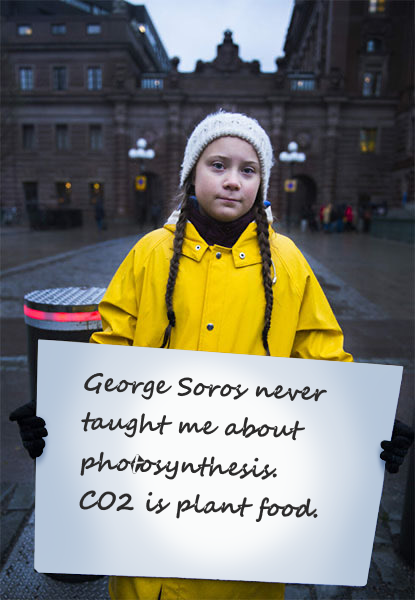}} \\
\subcaptionbox{Reddit Retrieval\label{c}}{\includegraphics[width = 1.2in, height = 1.2in]{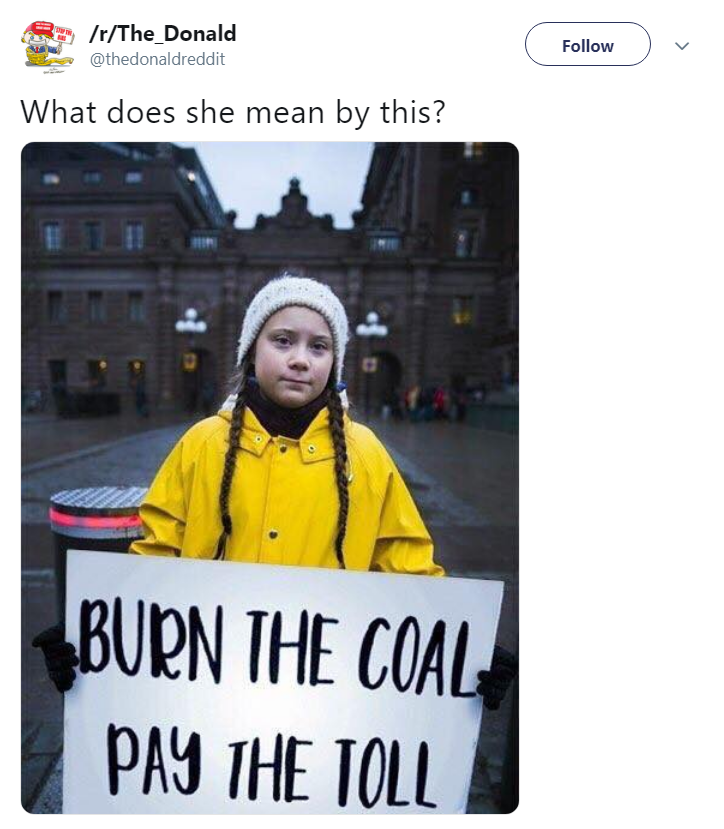}}&
\subcaptionbox{Reddit Retrieval\label{d}}{\includegraphics[width = 1.2in, height = 1.2in]{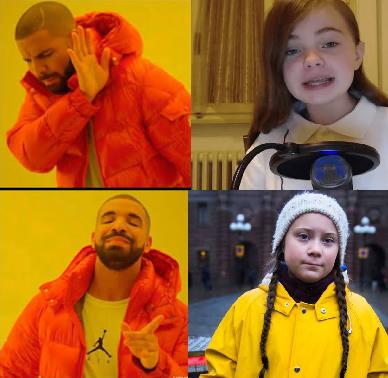}} \\
\end{tabular}
\caption{A 4chan query image (a) and three of its successfully retrieved matches from Reddit (b--d). 
Notice that these matches and manipulations go beyond traditional near-duplicate scenarios.}
\label{4chan_example}
\end{figure}

For this case study, we used two ``query'' datasets, from Twitter and 4chan, described below. For the training, we used $120$ images from each query dataset. We used each image to query an expanded Reddit index.
Among the $120$ Twitter queries we examined, in $28$ queries ($23.33\%$) the first retrieved result from Reddit was a duplicate or near-duplicate image; for 4chan  this number was $24$ ($20\%$).
Figure~\ref{4chan_example} shows an example of one 4chan query image and example of its successfully retrieved near-duplicates from the subreddits index. 


We used all queries and the first retrieved results, annotated as described above, resulting in 240 training examples. For each example we have two images (the query and the index match), use two features (the pHash distance and the retrieval similarity scores), and a binary label (matching or not).

We used a Logistic Regression classifier, after comparing several machine learning models\footnote{including SVM, Decision Tree, and Random Forest.} using leave-one-out cross-validation, and with various parameter configurations.
We validated the results on a held-out test set of additional 55 queries randomly selected from the Twitter and 4chan datasets, achieving accuracy of $100\%$ and AUC of~$1.0$ for the automatic classification of the first retrieved result. For these images, $10$ of $55$ were annotated as matching. Thus, the naive-majority-class accuracy, i.e. the accuracy when predicting always the major class (i.e., not matching), is $81.5\%$. Hence, our automatic classification process significantly improves the results. 
We also used this held-out test set to examine the generalization of the process for the classification of lower-scoring images (beyond the first result). For images retrieved at the second, third and fourth ranks, the process achieved accuracy of $92.8\%, 94.6\%, 96.4\%$ and AUC of $0.8, 0.75, 0.79$, respectively.
All these benchmarks outperform the naive-majority-class accuracy, which is $85.5\%, 89.1\% $ and $90.1\%$ for the second, third and fourth ranks, respectively.
We did not test the results of images retrieved for even lower scores since this would be a very different and much more difficult task than the task we trained the model for; future implementations could provide additional training to support this scenario. Furthermore, in this case study scenario, the images in these low ranks are mostly not matching results.

\subsubsection{Datasets.} 
For this case study we use two datasets as query sources: a political images dataset from Twitter, and a dataset of 4chan ``/pol/ - Politically Incorrect'' images. 
Our Twitter query image dataset is a random selection of $10,413$ tweets with images, selected from tweets published between September $2018$ to January $2019$ by candidates in the U.S. 2018 Congressional elections.
Our 4chan query images are $12,384$ images randomly selected from the ``/pol/ - Politically Incorrect'' 4chan dataset described in Section~\ref{dataset}.
We query the images by using our Reddit index, which contains $631,018$ images we crawled 
from the political subreddits listed in Section~\ref{dataset}. 

Note that these datasets consist of images naturally posted by users on social media, without any constraints. As such, they might include a variety of non-trivial manipulations and combinations of multiple manipulations, e.g., photoshopped images, collages, memes, etc.
This noisy set of images therefore better reflects potential real-world performance.

\subsubsection{Procedure.} 
We classified the first retrieved result of the $22,797$ Twitter and 4chan query images, by applying the retrieval and classification processes as outlined above.
We use post metadata to extract the date of posting for each image. We then had, for each positive match, the lag of time between their publication dates on the different social media platforms. 

Since we analyzed only the first retrieved result for each query image, the analysis of the lag of time was applied to one most visually-similar matching image that appears in the subreddits index.
Of course, a query image may have several near-duplicates in the index, with different dates of publication.
For this case study we simplify the formulations to present a process which can be used for a first-pass understanding of the dynamics.
Nevertheless, for full coverage of near-duplicates publications, the process can be repeated with additional queries and more than one retrieved result for each. 

\subsubsection{Results.}
Overall, $18.42\%$ of the images in the query datasets also appeared in the subreddits index
: $20.39\%$ ($2,123$) of the Twitter and
$16.76\%$ ($2,076$) of the 4chan query images. 

\begin{figure}[h]
\centering
\begin{tabular}{cc} 
{\includegraphics[width=1.0\columnwidth]{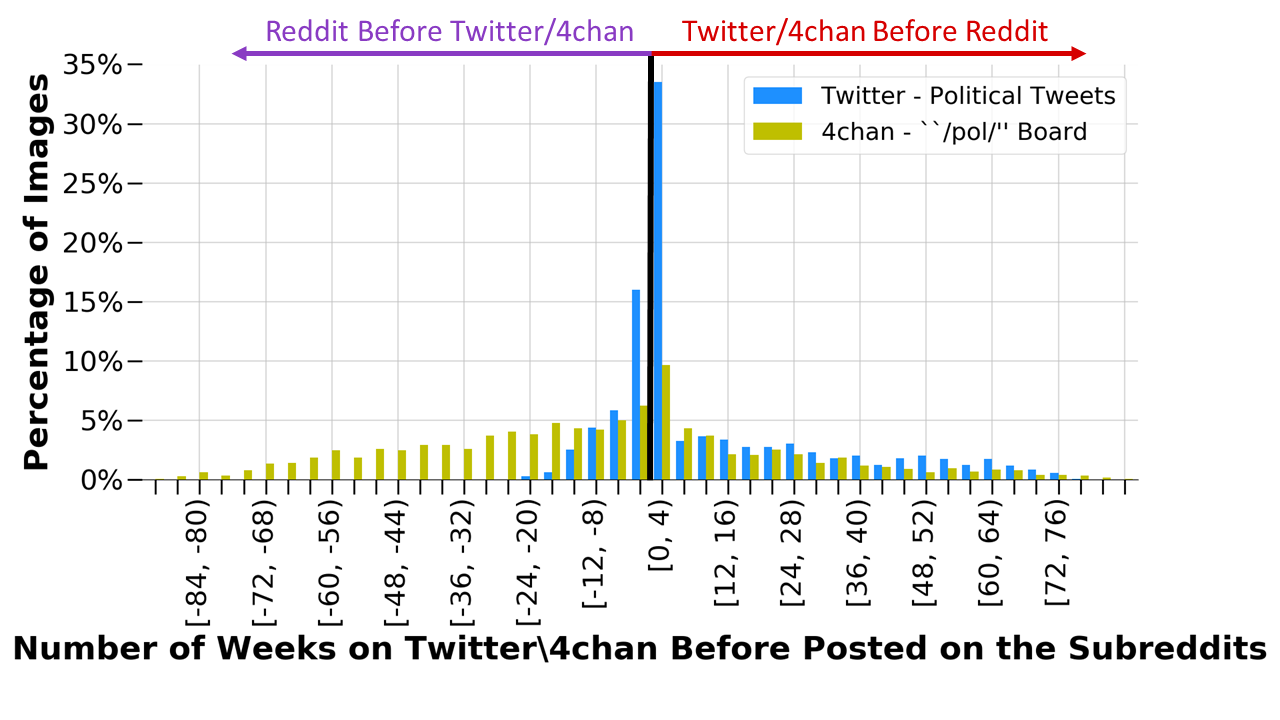}}
\end{tabular}
\vspace{-0.55cm}
\caption{Distribution of the percentage of Twitter/4chan images based on number of weeks before a matching image was posted on the subreddits. 
Positive values indicate that the publication on Twitter/4chan was prior the publication on the subreddits. 
}
\label{fig:combine_dates_histogram}
\end{figure}

For these matching images, we conducted analysis of their date of publication on Twitter/4chan, compared to the date of publication of their matching image in the subreddits index. 
Figure~\ref{fig:combine_dates_histogram} presents the percentage of the total number of images that were classified as visually-similar in each query set, based on the number of weeks that passed before (right of the~$0$ on the x-axis) or after (left) a matching image was posted in the subreddits index. 
The figure shows that, as expected, for both Twitter and 4chan, most images were in very close proximity to their Reddit matches.
For example, the tallest bars indicates that $33.5\%$ of Twitter images (blue) and $9.9\%$ of 4chan images (yellow) that matched subreddit images, were published between zero and three weeks before their subreddit match.
For 4chan, the distribution seems to be heavier on the left, meaning that in most of the cases images are posted on Reddit first, some times many weeks before.
In contrast, for Twitter we observed a heavier distribution on the right, which indicates that most of the matching images were first published on Twitter.
Note that the differences in the distributions is affected by the publication dates of the sampled query images.
A reasonable explanation may be the fact that the Twitter dataset covers images that were published on January 2019 at the latest, while 4chan and Reddit datasets lasted until March 2020.
That is, by using these examined datasets, the potential maximal number of weeks until the publication on Twitter (relatively to the publication date on the subreddits) is lower than for 4chan.
Nevertheless, the differences around the zero point show that this method can potentially expose different dynamics between platforms.

While a more robust and elaborate analysis remains for future work, this case study demonstrates the potential use of the process to automatically detect spread of visual content across social media platforms.

\section{Conclusions}
In this work, we evaluated the potential of several workflows for detecting image near-duplicates in the context of social media platforms. 

Our findings show that, in the social media context, several methods based on common techniques from the literature perform near-duplicates retrieval with reasonable accuracy. We have shown that one of these techniques performs well for a varied set of image tweaks and manipulations, though not for every such type. For example, the method was less robust to extreme manipulations like significant cropping of the image.
There are a number of improvements we can add to our indexing and retrieval process to improve accuracy even further~\cite{dong2012high,ke2004efficient,an2017near,chum2007scalable,zhu2008near,moreira2018image,pinto2017provenance}, including using a hybrid approach that builds on the strengths of the different computer vision methods.

Our contribution --- the methods and the dataset --- can assist the research community in performing various tasks related to tracking visual content distribution on social media platforms.
We make our code available to researchers upon request.
Moreover, these capabilities can support the development of systems to help individuals (e.g., journalists) to assess the veracity of visual content. 
Further, as we have shown in our case studies, the availability of such tool can support larger-scale research that builds on automated methods for tracking images through social media. 

\section{Acknowledgement}
This project was partially supported by a collaboration grant from the Jacobs Technion-Cornell Institute at Cornell Tech.

\bibliography{bib}

\begin{thebibliography}{}

\bibitem[\protect\citeauthoryear{Abdi and Williams}{2010}]{abdi2010principal}
Abdi, H., and Williams, L.~J.
\newblock 2010.
\newblock Principal component analysis.
\newblock {\em Wiley interdisciplinary reviews: computational statistics}
  2(4):433--459.

\bibitem[\protect\citeauthoryear{Allcott and
  Gentzkow}{2017}]{allcott2017social}
Allcott, H., and Gentzkow, M.
\newblock 2017.
\newblock Social media and fake news in the 2016 election.
\newblock {\em Journal of economic perspectives} 31(2):211--36.

\bibitem[\protect\citeauthoryear{An \bgroup et al\mbox.\egroup
  }{2017}]{an2017near}
An, S.; Huang, Z.; Chen, Y.; and Weng, D.
\newblock 2017.
\newblock Near duplicate product image detection based on binary hashing.
\newblock In {\em Proceedings of the 2017 international conference on deep
  learning technologies},  75--80.

\bibitem[\protect\citeauthoryear{Babenko \bgroup et al\mbox.\egroup
  }{2014}]{babenko2014neural}
Babenko, A.; Slesarev, A.; Chigorin, A.; and Lempitsky, V.
\newblock 2014.
\newblock Neural codes for image retrieval.
\newblock In {\em European conference on computer vision},  584--599.
\newblock Springer.

\bibitem[\protect\citeauthoryear{Barroso, Dean, and
  Holzle}{2003}]{barroso2003web}
Barroso, L.~A.; Dean, J.; and Holzle, U.
\newblock 2003.
\newblock Web search for a planet: The google cluster architecture.
\newblock {\em IEEE micro} 23(2):22--28.

\bibitem[\protect\citeauthoryear{Bromley \bgroup et al\mbox.\egroup
  }{1994}]{bromley1994signature}
Bromley, J.; Guyon, I.; LeCun, Y.; S{\"a}ckinger, E.; and Shah, R.
\newblock 1994.
\newblock Signature verification using a" siamese" time delay neural network.
\newblock In {\em Advances in neural information processing systems},
  737--744.

\bibitem[\protect\citeauthoryear{Chesney and Citron}{2019}]{chesney2019deep}
Chesney, B., and Citron, D.
\newblock 2019.
\newblock Deep fakes: A looming challenge for privacy, democracy, and national
  security.
\newblock {\em Calif. L. Rev.} 107:1753.

\bibitem[\protect\citeauthoryear{Chum \bgroup et al\mbox.\egroup
  }{2007}]{chum2007scalable}
Chum, O.; Philbin, J.; Isard, M.; and Zisserman, A.
\newblock 2007.
\newblock Scalable near identical image and shot detection.
\newblock In {\em Proceedings of the 6th ACM international conference on Image
  and video retrieval},  549--556.

\bibitem[\protect\citeauthoryear{Dalal and Triggs}{2005}]{dalal2005histograms}
Dalal, N., and Triggs, B.
\newblock 2005.
\newblock Histograms of oriented gradients for human detection.
\newblock In {\em 2005 IEEE computer society conference on computer vision and
  pattern recognition (CVPR'05)}, volume~1,  886--893.
\newblock IEEE.

\bibitem[\protect\citeauthoryear{Dang \bgroup et al\mbox.\egroup
  }{2017}]{dang2017offline}
Dang, A.; Moh’d, A.; Gruzd, A.; Milios, E.; and Minghim, R.
\newblock 2017.
\newblock An offline--online visual framework for clustering memes in social
  media.
\newblock In {\em From Social Data Mining and Analysis to Prediction and
  Community Detection}. Springer.
\newblock  1--29.

\bibitem[\protect\citeauthoryear{Dong \bgroup et al\mbox.\egroup
  }{2012}]{dong2012high}
Dong, W.; Wang, Z.; Charikar, M.; and Li, K.
\newblock 2012.
\newblock High-confidence near-duplicate image detection.
\newblock In {\em Proceedings of the 2nd acm international conference on
  multimedia retrieval},  1--8.

\bibitem[\protect\citeauthoryear{Dubey \bgroup et al\mbox.\egroup
  }{2018}]{dubey2018memesequencer}
Dubey, A.; Moro, E.; Cebrian, M.; and Rahwan, I.
\newblock 2018.
\newblock Memesequencer: Sparse matching for embedding image macros.
\newblock In {\em Proceedings of the 2018 World Wide Web Conference},
  1225--1235.

\bibitem[\protect\citeauthoryear{Ge \bgroup et al\mbox.\egroup
  }{2013}]{ge2013optimized}
Ge, T.; He, K.; Ke, Q.; and Sun, J.
\newblock 2013.
\newblock Optimized product quantization for approximate nearest neighbor
  search.
\newblock In {\em Proceedings of the IEEE Conference on Computer Vision and
  Pattern Recognition},  2946--2953.

\bibitem[\protect\citeauthoryear{Grinberg \bgroup et al\mbox.\egroup
  }{2019}]{grinberg2019fake}
Grinberg, N.; Joseph, K.; Friedland, L.; Swire-Thompson, B.; and Lazer, D.
\newblock 2019.
\newblock Fake news on twitter during the 2016 us presidential election.
\newblock {\em Science} 363(6425):374--378.

\bibitem[\protect\citeauthoryear{Gu{\'e}rin \bgroup et al\mbox.\egroup
  }{2017}]{guerin2017cnn}
Gu{\'e}rin, J.; Gibaru, O.; Thiery, S.; and Nyiri, E.
\newblock 2017.
\newblock Cnn features are also great at unsupervised classification.
\newblock {\em arXiv preprint arXiv:1707.01700}.

\bibitem[\protect\citeauthoryear{Guess, Nagler, and
  Tucker}{2019}]{guess2019less}
Guess, A.; Nagler, J.; and Tucker, J.
\newblock 2019.
\newblock Less than you think: Prevalence and predictors of fake news
  dissemination on facebook.
\newblock {\em Science advances} 5(1):eaau4586.

\bibitem[\protect\citeauthoryear{Gupta \bgroup et al\mbox.\egroup
  }{2013}]{gupta2013faking}
Gupta, A.; Lamba, H.; Kumaraguru, P.; and Joshi, A.
\newblock 2013.
\newblock Faking sandy: characterizing and identifying fake images on twitter
  during hurricane sandy.
\newblock In {\em Proceedings of the 22nd international conference on World
  Wide Web},  729--736.

\bibitem[\protect\citeauthoryear{He, Wen, and Sun}{2013}]{he2013k}
He, K.; Wen, F.; and Sun, J.
\newblock 2013.
\newblock K-means hashing: An affinity-preserving quantization method for
  learning binary compact codes.
\newblock In {\em Proceedings of the IEEE conference on computer vision and
  pattern recognition},  2938--2945.

\bibitem[\protect\citeauthoryear{Highfield and
  Leaver}{2016}]{highfield2016instagrammatics}
Highfield, T., and Leaver, T.
\newblock 2016.
\newblock Instagrammatics and digital methods: Studying visual social media,
  from selfies and gifs to memes and emoji.
\newblock {\em Communication Research and Practice} 2(1):47--62.

\bibitem[\protect\citeauthoryear{Huh \bgroup et al\mbox.\egroup
  }{2018}]{huh2018fighting}
Huh, M.; Liu, A.; Owens, A.; and Efros, A.~A.
\newblock 2018.
\newblock Fighting fake news: Image splice detection via learned
  self-consistency.
\newblock In {\em Proceedings of the European Conference on Computer Vision
  (ECCV)},  101--117.

\bibitem[\protect\citeauthoryear{Johnson, Douze, and
  J{\'e}gou}{2019}]{johnson2019billion}
Johnson, J.; Douze, M.; and J{\'e}gou, H.
\newblock 2019.
\newblock Billion-scale similarity search with gpus.
\newblock {\em IEEE Transactions on Big Data}.

\bibitem[\protect\citeauthoryear{Kanungo \bgroup et al\mbox.\egroup
  }{2002}]{kanungo2002efficient}
Kanungo, T.; Mount, D.~M.; Netanyahu, N.~S.; Piatko, C.~D.; Silverman, R.; and
  Wu, A.~Y.
\newblock 2002.
\newblock An efficient k-means clustering algorithm: Analysis and
  implementation.
\newblock {\em IEEE transactions on pattern analysis and machine intelligence}
  24(7):881--892.

\bibitem[\protect\citeauthoryear{Ke and Sukthankar}{2004}]{ke2004pca}
Ke, Y., and Sukthankar, R.
\newblock 2004.
\newblock Pca-sift: A more distinctive representation for local image
  descriptors.
\newblock In {\em Proceedings of the 2004 IEEE Computer Society Conference on
  Computer Vision and Pattern Recognition, 2004. CVPR 2004.}, volume~2,
  II--II.
\newblock IEEE.

\bibitem[\protect\citeauthoryear{Ke \bgroup et al\mbox.\egroup
  }{2004}]{ke2004efficient}
Ke, Y.; Sukthankar, R.; Huston, L.; Ke, Y.; and Sukthankar, R.
\newblock 2004.
\newblock Efficient near-duplicate detection and sub-image retrieval.
\newblock In {\em Acm Multimedia}, volume~4, ~5.
\newblock Citeseer.

\bibitem[\protect\citeauthoryear{Lago, Phan, and Boato}{2018}]{lago2018image}
Lago, F.; Phan, Q.-T.; and Boato, G.
\newblock 2018.
\newblock Image forensics in online news.
\newblock In {\em 2018 IEEE 20th International Workshop on Multimedia Signal
  Processing (MMSP)},  1--6.
\newblock IEEE.

\bibitem[\protect\citeauthoryear{Lago, Phan, and Boato}{2019}]{lago2019visual}
Lago, F.; Phan, Q.-T.; and Boato, G.
\newblock 2019.
\newblock Visual and textual analysis for image trustworthiness assessment
  within online news.
\newblock {\em Security and Communication Networks} 2019.

\bibitem[\protect\citeauthoryear{Lloyd}{1982}]{lloyd1982least}
Lloyd, S.
\newblock 1982.
\newblock Least squares quantization in pcm.
\newblock {\em IEEE transactions on information theory} 28(2):129--137.

\bibitem[\protect\citeauthoryear{Lowe}{2004}]{lowe2004distinctive}
Lowe, D.~G.
\newblock 2004.
\newblock Distinctive image features from scale-invariant keypoints.
\newblock {\em International journal of computer vision} 60(2):91--110.

\bibitem[\protect\citeauthoryear{Marra \bgroup et al\mbox.\egroup
  }{2018}]{marra2018detection}
Marra, F.; Gragnaniello, D.; Cozzolino, D.; and Verdoliva, L.
\newblock 2018.
\newblock Detection of gan-generated fake images over social networks.
\newblock In {\em 2018 IEEE Conference on Multimedia Information Processing and
  Retrieval (MIPR)},  384--389.
\newblock IEEE.

\bibitem[\protect\citeauthoryear{Marra \bgroup et al\mbox.\egroup
  }{2019}]{marra2019gans}
Marra, F.; Gragnaniello, D.; Verdoliva, L.; and Poggi, G.
\newblock 2019.
\newblock Do gans leave artificial fingerprints?
\newblock In {\em 2019 IEEE Conference on Multimedia Information Processing and
  Retrieval (MIPR)},  506--511.
\newblock IEEE.

\bibitem[\protect\citeauthoryear{Marwick and Lewis}{2017}]{marwick2017media}
Marwick, A., and Lewis, R.
\newblock 2017.
\newblock Media manipulation and disinformation online.
\newblock {\em New York: Data \& Society Research Institute}.

\bibitem[\protect\citeauthoryear{Matatov \bgroup et al\mbox.\egroup
  }{2019}]{matatovdejavu}
Matatov, H.; Bechhofer, A.; Aroyo, L.; Amir, O.; and Naaman, M.
\newblock 2019.
\newblock {DejaVu}: A system for journalists to collaboratively address visual
  misinformation.
\newblock In {\em Proceedings of Computation+Journalism Symposium}.

\bibitem[\protect\citeauthoryear{McCloskey and
  Albright}{2018}]{mccloskey2018detecting}
McCloskey, S., and Albright, M.
\newblock 2018.
\newblock Detecting gan-generated imagery using color cues.
\newblock {\em arXiv preprint arXiv:1812.08247}.

\bibitem[\protect\citeauthoryear{Monga and Evans}{2006}]{monga2006perceptual}
Monga, V., and Evans, B.~L.
\newblock 2006.
\newblock Perceptual image hashing via feature points: performance evaluation
  and tradeoffs.
\newblock {\em IEEE transactions on Image Processing} 15(11):3452--3465.

\bibitem[\protect\citeauthoryear{Moreira \bgroup et al\mbox.\egroup
  }{2018}]{moreira2018image}
Moreira, D.; Bharati, A.; Brogan, J.; Pinto, A.; Parowski, M.; Bowyer, K.~W.;
  Flynn, P.~J.; Rocha, A.; and Scheirer, W.~J.
\newblock 2018.
\newblock Image provenance analysis at scale.
\newblock {\em IEEE Transactions on Image Processing} 27(12):6109--6123.

\bibitem[\protect\citeauthoryear{Paris and Donovan}{2019}]{paris2019deepfakes}
Paris, B., and Donovan, J.
\newblock 2019.
\newblock Deepfakes and cheap fakes.
\newblock {\em United States of America: Data \& Society}.

\bibitem[\protect\citeauthoryear{Pinto \bgroup et al\mbox.\egroup
  }{2017}]{pinto2017provenance}
Pinto, A.; Moreira, D.; Bharati, A.; Brogan, J.; Bowyer, K.; Flynn, P.;
  Scheirer, W.; and Rocha, A.
\newblock 2017.
\newblock Provenance filtering for multimedia phylogeny.
\newblock In {\em 2017 IEEE International Conference on Image Processing
  (ICIP)},  1502--1506.
\newblock IEEE.

\bibitem[\protect\citeauthoryear{Radenovi{\'c} \bgroup et al\mbox.\egroup
  }{2018}]{radenovic2018revisiting}
Radenovi{\'c}, F.; Iscen, A.; Tolias, G.; Avrithis, Y.; and Chum, O.
\newblock 2018.
\newblock Revisiting oxford and paris: Large-scale image retrieval
  benchmarking.
\newblock In {\em Proceedings of the IEEE Conference on Computer Vision and
  Pattern Recognition},  5706--5715.

\bibitem[\protect\citeauthoryear{Ramisa \bgroup et al\mbox.\egroup
  }{2017}]{ramisa2017breakingnews}
Ramisa, A.; Yan, F.; Moreno-Noguer, F.; and Mikolajczyk, K.
\newblock 2017.
\newblock Breakingnews: Article annotation by image and text processing.
\newblock {\em IEEE transactions on pattern analysis and machine intelligence}
  40(5):1072--1085.

\bibitem[\protect\citeauthoryear{Rublee \bgroup et al\mbox.\egroup
  }{2011}]{rublee2011orb}
Rublee, E.; Rabaud, V.; Konolige, K.; and Bradski, G.
\newblock 2011.
\newblock Orb: An efficient alternative to sift or surf.
\newblock In {\em 2011 International conference on computer vision},
  2564--2571.
\newblock Ieee.

\bibitem[\protect\citeauthoryear{Sermanet \bgroup et al\mbox.\egroup
  }{2013}]{sermanet2013overfeat}
Sermanet, P.; Eigen, D.; Zhang, X.; Mathieu, M.; Fergus, R.; and LeCun, Y.
\newblock 2013.
\newblock Overfeat: Integrated recognition, localization and detection using
  convolutional networks.
\newblock {\em arXiv preprint arXiv:1312.6229}.

\bibitem[\protect\citeauthoryear{Shu \bgroup et al\mbox.\egroup
  }{2017}]{shu2017fake}
Shu, K.; Sliva, A.; Wang, S.; Tang, J.; and Liu, H.
\newblock 2017.
\newblock Fake news detection on social media: A data mining perspective.
\newblock {\em ACM SIGKDD Explorations Newsletter} 19(1):22--36.

\bibitem[\protect\citeauthoryear{Simonyan and
  Zisserman}{2014}]{simonyan2014very}
Simonyan, K., and Zisserman, A.
\newblock 2014.
\newblock Very deep convolutional networks for large-scale image recognition.
\newblock {\em arXiv preprint arXiv:1409.1556}.

\bibitem[\protect\citeauthoryear{Vattani}{2011}]{vattani2011k}
Vattani, A.
\newblock 2011.
\newblock K-means requires exponentially many iterations even in the plane.
\newblock {\em Discrete \& Computational Geometry} 45(4):596--616.

\bibitem[\protect\citeauthoryear{Vosoughi, Roy, and
  Aral}{2018}]{vosoughi2018spread}
Vosoughi, S.; Roy, D.; and Aral, S.
\newblock 2018.
\newblock The spread of true and false news online.
\newblock {\em Science} 359(6380):1146--1151.

\bibitem[\protect\citeauthoryear{Xuan \bgroup et al\mbox.\egroup
  }{2019}]{xuan2019generalization}
Xuan, X.; Peng, B.; Wang, W.; and Dong, J.
\newblock 2019.
\newblock On the generalization of gan image forensics.
\newblock In {\em Chinese Conference on Biometric Recognition},  134--141.
\newblock Springer.

\bibitem[\protect\citeauthoryear{Yang \bgroup et al\mbox.\egroup
  }{2017}]{yang2017yum}
Yang, L.; Hsieh, C.-K.; Yang, H.; Pollak, J.~P.; Dell, N.; Belongie, S.; Cole,
  C.; and Estrin, D.
\newblock 2017.
\newblock Yum-me: a personalized nutrient-based meal recommender system.
\newblock {\em ACM Transactions on Information Systems (TOIS)} 36(1):1--31.

\bibitem[\protect\citeauthoryear{Zannettou \bgroup et al\mbox.\egroup
  }{2018}]{zannettou2018origins}
Zannettou, S.; Caulfield, T.; Blackburn, J.; De~Cristofaro, E.; Sirivianos, M.;
  Stringhini, G.; and Suarez-Tangil, G.
\newblock 2018.
\newblock On the origins of memes by means of fringe web communities.
\newblock In {\em Proceedings of the Internet Measurement Conference 2018},
  188--202.

\bibitem[\protect\citeauthoryear{Zhang, Karaman, and
  Chang}{2019}]{zhang2019detecting}
Zhang, X.; Karaman, S.; and Chang, S.-F.
\newblock 2019.
\newblock Detecting and simulating artifacts in gan fake images.
\newblock {\em arXiv preprint arXiv:1907.06515}.

\bibitem[\protect\citeauthoryear{Zhu \bgroup et al\mbox.\egroup
  }{2008}]{zhu2008near}
Zhu, J.; Hoi, S.~C.; Lyu, M.~R.; and Yan, S.
\newblock 2008.
\newblock Near-duplicate keyframe retrieval by nonrigid image matching.
\newblock In {\em Proceedings of the 16th ACM international conference on
  Multimedia},  41--50.

\end{thebibliography}
\bibliographystyle{aaai}

\end{document}